# Raman Signatures of Polytypism in Molybdenum Disulfide

*Jae-Ung Lee,[†] Kangwon Kim,[†] Songhee Han,[†] Gyeong Hee Ryu,[‡] Zonghoon Lee,[‡]*
*and Hyeonsik Cheong[†,*]*

[†]Department of Physics, Sogang University, Seoul 04107, Korea

[‡]School of Materials Science and Engineering, Ulsan National Institute of Science and Technology (UNIST), Ulsan 44919, Korea

ABSTRACT

Since the stacking order sensitively affects various physical properties of layered materials, accurate determination of the stacking order is important for studying the basic properties of these materials as well as for device applications. Because 2H-molybdenum disulfide (MoS$_2$) is most common in nature, most studies so far have focused on 2H-MoS$_2$. However, we found that the 2H, 3R, and mixed stacking sequences exist in few-layer MoS$_2$ exfoliated from natural molybdenite crystals. The crystal structures are confirmed by HR-TEM measurements. The Raman signatures of different polytypes are investigated by using 3 different excitation energies which are non-resonant and resonant with A and C excitons, respectively. The low-frequency breathing and shear modes show distinct differences for each polytype whereas the high-frequency intra-layer modes show little difference. For resonant excitations at 1.96 and 2.81 eV, distinct features are observed which enable determination of the stacking order.

Polytypism, a special type of polymorphism in layered materials, refers to different stacking sequences of monolayers with the same structure.[1–3] Since stacking sequence is one of the key attributes of layered materials, the effects of different stacking sequences on the electronic and other properties of 2-dimensional (2D) materials are of great interest. Although the crystal structure of each layer is identical, the properties of few-layer crystals are sensitively dependent on the stacking sequence due to different inter-layer interactions. For example, in the case of graphene, two stable stacking orders, ABA (Bernal) and ABC (rhombohedral) stacking orders, predominantly exist in nature.[4,5] Several studies have revealed the influence of the stacking order on transport[6,7] and optical properties.[8–12] Hence, identifying the stacking sequences has become an important issue, and Raman spectroscopy has proven to be a reliable and easy characterization tool to identify stacking orders in few-layer graphene.[9–11,13] Since the Raman spectrum reflects the phonon dispersion and the electronic band structure, it is an ideal tool for fingerprinting the polytypes of graphene without complicated sample preparations.

In layered transition metal dichalcogenides (TMDs), polymorphism and polytypism are more important than in graphene because of more complex crystal structures. Among TMD materials, molybdenum disulfide ($MoS_2$) is the most extensively studied. Thanks to a finite bandgap, electronic applications such as field effect transistors[14,15] or photodetectors[16,17] are explored. Single-layer $MoS_2$ comprises a monolayer of Mo atoms sandwiched between two sulfur layers, forming a 'trilayer' (TL).[18,19] Each TL is connected *via* weak van der Waals interactions. Like other 2D materials, $MoS_2$ has several polymorphs. For single-layer $MoS_2$, there are two types of polymorphs: trigonal prism (1H-$MoS_2$) and octahedral coordination (1T-$MoS_2$). Since 1T-$MoS_2$ is metastable, only the trigonal phase is found in natural bulk $MoS_2$.[19] In few-layer $MoS_2$, stacking of 1H-$MoS_2$ layers results in two polytypes: hexagonal 2H and rhombohedral 3R.

Naturally occurring MoS$_2$ crystals (molybdenite) are predominantly 2H type due to its lower formation energy.[19,20] Because of the rarity of 3R-MoS$_2$, most of studies have focused on 2H-MoS$_2$ and only few studies have been done for 3R-MoS$_2$.[21,22] Raman spectroscopy has become an important characterization tool for MoS$_2$ as well, and the Raman signatures of few-layer MoS$_2$[18,23–25] and the effect of resonance with exciton states have been extensively studied.[25–28] In all these studies, only 2H-MoS$_2$ was investigated. In this work, we identified few-layer 3R-MoS$_2$ among samples obtained from natural molybdenite crystals through high resolution transmission electron microscopy (HR-TEM) and compared the Raman spectra of 2H-, 3R- and mixed MoS$_2$. By using 3 different excitation energies, the resonance effects were also compared for different polytypes.

**RESULTS AND DISCUSSIONS**

The crystal structures of 2H- and 3R-MoS$_2$ are shown in Fig. 1 schematically. In 2H stacking order, each layer is rotated 180° with respect to the adjacent layer as shown in Fig. 1(a). In contrast, each layer in 3R stacking order has the same orientation as shown in Fig. 1(b). Figures 1(c) and (d) show the top-view of 2H- and 3R-MoS$_2$. The lattice is formed in 2H-MoS$_2$ without any atom at the center of a hexagon, whereas in 3R-MoS$_2$ an atom is located at the center of a hexagon. For 3R stacking, there are several variations according to different relative shift of adjacent layers, whereas there is only one type of 2H for each number of layers. For 2TL MoS$_2$, for example, only one type of 3R stacking is possible but for 3TL MoS$_2$, four inequivalent 3R stacking orders can exist as shown in Fig. S1.[2] In addition, a mixed stacking of 2H-3R or 3R-2H can exist (2H-3R and 3R-2H structures are equivalent). Figures 1(b) and (d) show the 3R-3 type structure. The space groups and Raman tensors of different polytypes of 3TL MoS$_2$ are summarized in Table S1. In Raman measurements on several few-layer MoS$_2$ samples, we encountered cases where two samples with the same thickness as measured by atomic force microscopy show clearly different Raman spectra. In order to understand these differences, we carried out HR-TEM measurements and found that the samples have different stacking orders. Figures 1(e) and (f) and S2(a) and (b) show HR-TEM images of 3TL MoS$_2$ in 2H and 3R stacking orders. These images can be correlated with the crystal structure shown in Figs. 1(c) and (d), which determines the stacking orders to be 2H and 3R, respectively. Figure S3 shows optical microscope images of some samples that contain the 3R phase.

Figure 2 shows the Raman spectra of 2H and 3R MoS$_2$ (3TL) measured with the 2.41 eV (514.5 nm) excitation energy. For 2H, the 2.41 eV excitation energy can be regarded as non-resonant.[26] Although different stacking orders would modify the electronic band structure, the exciton state energies do not seem to be affected much as one can see from the photoluminescence (PL) data in Fig. S4: the PL spectra are fairly similar. The angle resolved photoemission[21] and optical transmission[29] results from bulk MoS$_2$ also have shown that the overall band structure is similar for 2H and 3R polytypes, except for the spin polarizations.[21] Therefore, we consider 2.41 eV as non-resonant for both polytypes. The main $E_{2g}^1$ and $A_{1g}$ modes are observed at ~383 and ~407 cm$^{-1}$, respectively, for both polytypes. These modes correspond to the intra-layer vibrations along the in-plane or out-of-plane directions, respectively. There is no significant difference in the intra-layer vibration modes, which implies that each constituent layer is almost identical for both polytypes. This result is consistent with previous Raman results on synthetic 3R-MoS$_2$.[21] The possibility of the 1T phase can be ruled out since the Raman peak position near the main modes should be different for 1T-MoS$_2$.[30–32] In the low-frequency region, shear and breathing modes, inter-layer in-plane and out-of-plane vibrations, respectively, are observed.[18,24,26] One can see that the shear mode frequency has little dependence on the stacking order. On the other hand, the breathing mode shows a clear difference: in 3R-MoS$_2$, the shear and breathing modes are clearly resolved whereas they overlap and are not resolved for 2H-MoS$_2$. The peak position of the breathing mode in 3R-MoS$_2$ is redshifted by ~3 cm$^{-1}$ from that of 2H-MoS$_2$. Our results suggest that the out-of-plane Young's modulus is smaller for the 3R polytype and the shear modulus does not depend much on the stacking order.

Although the Raman spectrum of 3R-MoS$_2$ has not been studied much, the Raman spectra of 3R-MoSe$_2$ and 3R-WSe$_2$ grown by chemical vapor deposition (CVD) have been studied.[2,3] It was found that the shear modes are insensitive to the stacking order, whereas the breathing mode of 3R-WSe$_2$ showed a trend similar to ours. The breathing modes of MoSe$_2$ were not detected in these studies. For 3TL-MoSe$_2$, 2H, 2H-3R mixed, and two variations of 3R stacking orders were identified by Raman spectroscopy.[2] In 3R-1 or 3R-2 types, the high frequency shear mode is weak and the low-frequency mode is allowed in backscattering whereas the low-frequency shear mode is forbidden in backscattering in 3R-3 or 3R-4 types. By comparing these results with ours, we can conclude that our 3TL-MoS$_2$ sample belongs to either 3R-3 or 3R-4 type in Fig. S1. These two variations cannot be distinguished by HR-TEM or Raman spectroscopy. In addition, we found a sample with areas that show low-frequency spectra that match the spectrum of neither 2H nor 3R polytypes. We interpret that this sample contains areas with mixed stacking of 2H and 3R polytypes. Figure 2(a) shows a spectrum from mixed (2H-3R) 3TL MoS$_2$, which is characterized by a breathing mode between those of 2H and 3R polytypes and a weak extra shear mode at ~16.5 cm$^{-1}$. These are similar to what was observed in mixed 3TL MoSe$_2$.[2] It is not likely that these samples correspond to mis-oriented (twisted) stacking. In samples of layered materials prepared by mechanical exfoliation, twisted stacking orders are not found except for the case of folded samples. Furthermore, the edge of the mixed phase area and that of the 3R area in Fig. S3 make 60°, which suggests that the layers are properly stacked. The optical microscope images in Fig. S3 show both 3R and mixed-phase samples. There is no clear difference in the optical contrast.

Resonance Raman scattering in MoS$_2$ shows some interesting effects.[25–28,33] For the 1.96 eV excitation energy, which is close to the A (or B) exciton states, several new features emerge. In the low-frequency region, a broad peak centered at the Rayleigh scattered laser line appears due to the resonance excitation of exciton states mediated by acoustic phonon scattering.[26] In addition, several new peaks and many second order Raman peaks are enhanced due to resonance with exciton or exciton-polariton states.[26] At the excitation energy of 2.81 eV, which is close to the C exciton state (~ 2.7 eV),[28,34,35] enhancement of some forbidden Raman modes $E_{1g}$ and $A_{2u}$ are reported.[26,27] Figure 3 shows the Raman spectra of the same set of 3TL MoS$_2$ samples measured with 1.96 eV (632.8 nm) and 2.81 eV (441.6 nm) excitation energies. For the 1.96 eV excitation, the peak positions of the main $E^1_{2g}$ and $A_{1g}$ modes as well as resonantly enhanced peaks are almost identical for 2H- and 3R-MoS$_2$ [Fig. 3(a)]. The relative intensities of the central peak and 'peak X'[26] are slightly smaller for 3R-MoS$_2$. The relative intensity of the 2LA peak with respect to the peak 'd' is smaller in 3R-MoS$_2$ than in 2H-MoS$_2$. This confirms that the exciton states are only slightly different for the two polytypes. However, for the mixed phase, the difference is more obvious. The peak 'b' is relatively enhanced with respect to the main $A_{1g}$ mode; and the peak 'c' is slightly red-shifted. Furthermore, a new peak at 54 cm$^{-1}$ labelled by '▲' appears only in the mixed phase. For 2H-MoS$_2$, the frequency of the second lowest breathing mode is ~50 cm$^{-1}$ in the diatomic chain model (DCM),[24] and we interpret that this peak

'▲' corresponds to the second breathing mode of mixed MoS$_2$. Figure 3(b) compares the Raman spectra of the same set of samples measured with the excitation energy of 2.81 eV. At this excitation energy, the breathing modes are relatively enhanced,[36] and so the difference between different stacking orders is more obvious. The selective enhancement of the in-plane $E_{2g}^1$ mode with respect to the out-of-plane $A_{1g}$ mode is observed when the excitation energy matches the C exciton state due to symmetry dependent exciton-phonon coupling.[28] A similar mechanism is suspected for the relative enhancement of the breathing mode at this excitation energy. Further study is needed to understand this phenomenon. For mixed MoS$_2$, the second shear mode at 16.5 cm$^{-1}$ is also relatively enhanced.

We measured several samples with thickness from 4TL to 7TL. Different stacking orders were identified based on the Raman spectrum. The overall trends are similar to the case of 3TL MoS$_2$ with the excitation energy of 1.96 eV as shown in Fig. 4. The spectra for 2H and 3R phases are very similar for each thickness. The inter-layer breathing mode, which shows the most obvious difference, is not well resolved at this excitation energy. On the other hand, several differences are observed in the mixed phase: 1) a relatively strong central peak and 'peak X'[26] are observed regardless of thickness; 2) the peak 'a' is relatively strong; 3) the peak 'b' is stronger than the main $A_{1g}$ mode; 4) the peak 'c' is relatively redshifted; 5) the 2LA peak is stronger relative to the peak 'd'; and 6) peaks labelled by '▲' appear. The peak 'b' might be a satellite of the $A_{1g}$ mode due to the Davydov splitting,[37–40] which should be sensitive to stacking. The peaks labelled by '▲' are redshifted as the number of layer increases, and additional peaks appear in thicker samples. These trends are consistent with the previous assignments of these peaks as higher-order breathing modes.

For the 2.41 eV and 2.81 eV excitation energies, the differences are obvious in the low-frequency range. The most important difference between 3R-MoS$_2$ and 2H-MoS$_2$ is the relative redshift of the lowest breathing mode for 3R-MoS$_2$. With the 2.81 eV excitation, the breathing modes are relatively enhanced for all thicknesses [Fig. 5(b)], and so the difference between the two polytypes is more readily identifiable. For all thicknesses, many additional low-frequency peaks are observed in mixed phase MoS$_2$. The positions of the observed shear, breathing, $E_{2g}^1$, and $A_{1g}$ modes are summarized in Fig. S5. In Fig. S6, the Raman spectra of different types of samples taken with the 2.41 eV excitation are compiled for 2 to 8TL. The polytypes of the samples presented in Figs. 2-5 are identified. For 3TL or thicker, several variations of 3R as well as mixed (2H and 3R) stacking orders are possible as mentioned earlier.[2] This makes it difficult to distinguish 3R and mixed phase for thicker samples. In Fig. S6, these types are collectively labelled as 'non-2H'. The spectra identified as 'mixed' are from the same area of the sample where mixed 3TL-MoS$_2$ was found.

In our study, we found a significant number of 3R or mixed phase samples: about 90% or more of our exfoliated few-layer MoS$_2$ samples turn out to be the 2H type, with the remainder belonging to either the 3R or the mixed phase. This suggests that the possibility of non-2H MoS$_2$ cannot be ignored even in natural molybdenite crystals. The small difference in the formation energy of 3R and 2H phases would explain the occurrence of 3R or mixed phase MoS$_2$ in natural as well as synthetic MoS$_2$. In some cases, the effect of different stacking order is significant. For example, the inversion symmetry is always broken for 3R-MoS$_2$, whereas 2H-MoS$_2$ has broken inversion symmetry only for odd-number TLs. Because the inversion symmetry plays an important role in valley physics and piezoelectricity of MoS$_2$, the possibility of polytypism needs to be carefully accounted for. Our results further demonstrate that the stacking order cannot be identified by simply measuring the peak frequency difference between the $E_{2g}^1$ and $A_{1g}$ modes or comparing the PL spectra. Careful comparison of Raman fingerprints of different polytypes, measured with different lasers would be necessary.

**CONCLUSIONS**

In conclusion, we found the 3R and mixed phases of $MoS_2$ among few-layer flakes obtained from natural molybdenite. For different stacking orders, the intra-layer vibrational modes and PL spectra are almost identical. For the inter-layer modes, the lowest breathing mode of 3R-$MoS_2$ is redshifted by ~3 cm$^{-1}$ with respect to that of 2H-$MoS_2$, which suggests that the inter-layer interaction is weaker for the 3R phase. For mixed phase $MoS_2$, higher order breathing modes and extra shear modes are observed. Comparison of such Raman fingerprints would facilitate determination of the polytypes of few-layer $MoS_2$.

**METHODS**

**Raman Measurements.** The samples were prepared on $SiO_2$/Si substrates by mechanical exfoliation from single-crystal bulk $MoS_2$ flakes (SPI supplies) from Otter Lake in Ontario, Canada, which is known to produce predominantly 2H-$MoS_2$.[41] The number of TLs was determined by combining optical contrast, Raman, photoluminescence (PL) and atomic force microscopy (AFM) measurements.[21,23,42] The 441.6-nm (2.81 eV) line of a He-Cd laser, the 514.5-nm (2.41 eV) line of an Ar ion laser, and the 632.8-nm (1.96 eV) line of a He-Ne laser were used as excitation sources. The laser beam was focused onto a sample by a 50× microscope objective lens (0.8 N.A.), and the scattered light was collected and collimated by the same objective. The scattered signal was dispersed by a Jobin-Yvon Horiba iHR550 spectrometer (2400 grooves/mm) and detected with a liquid-nitrogen-cooled back-illuminated charge-coupled-device (CCD) detector. To access the low-frequency range below 100 cm$^{-1}$, volume holographic filters (Ondax for 514.5-nm excitation, OptiGrate for 441.6 and 632.8-nm excitations) were used to clean the laser lines and reject the Rayleigh-scattered light. The laser power was kept below 50 μW in order to avoid local heating.

**HR-TEM Measurements.** MoS$_2$ flakes on the SiO$_2$/Si substrate were transferred onto a TEM grid.[43] The samples were analyzed using an aberration-corrected FEI Titan Cubed TEM (FEI Titan$^3$ G2 60-300), which was operated at 80 kV acceleration voltage with a monochromator. The microscope provides sub-Angstrom resolution at 80 kV and –13 ± 0.5 μm of spherical aberration (Cs). The atomic images were taken using a white atom contrast in order to obtain actual atom positions under properly focused conditions needed for direct image interpretation.

FIGURES

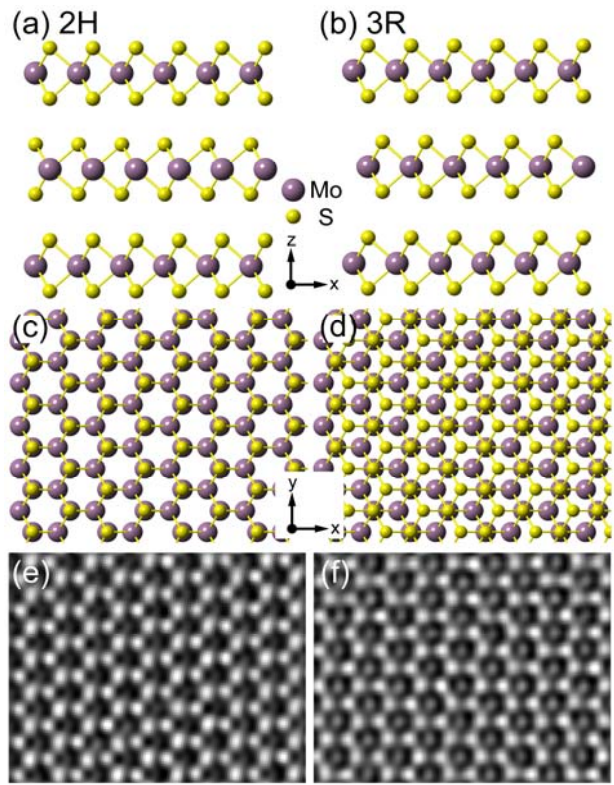

**Figure 1.** Crystal structures of 2H- and 3R-MoS$_2$. Side view of (a) 2H- and (b) 3R-MoS$_2$. Top view of (c) 2H- and (d) 3R-MoS$_2$. HR-TEM images of 3TL MoS$_2$ with (e) 2H and (f) 3R stacking orders.

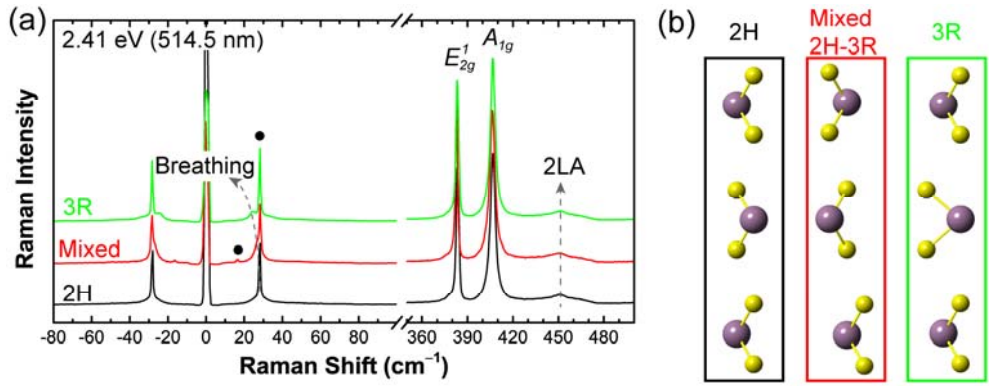

**Figure 2.** (a) Raman spectra of 3TL MoS$_2$ samples with different stacking orders, measured with the 2.41 eV (514.5 nm) excitation energy. Small circles (●) indicate the shear modes. (b) 2H (black), 3R (green), and mixed (red) stacking orders are shown with schematic side view of the crystal structures.

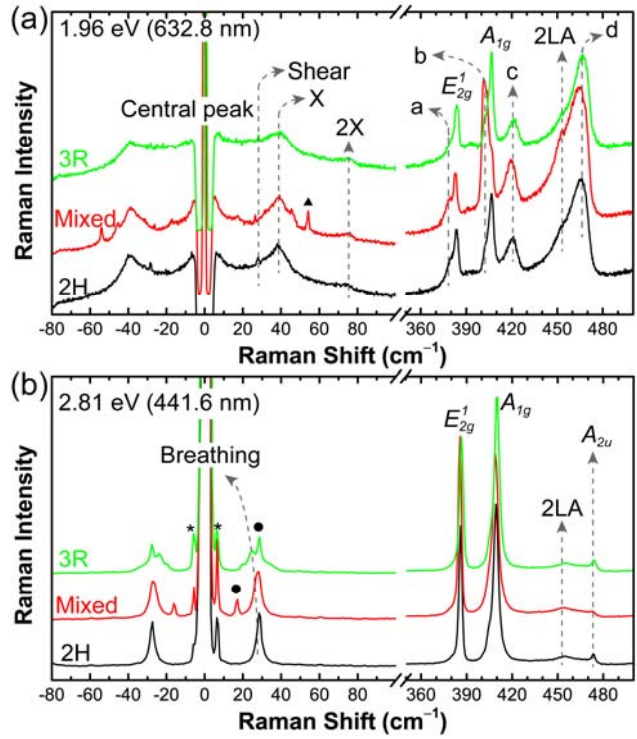

**Figure 3.** Raman spectra of 3TL MoS$_2$ samples with different stacking orders, measured with the (a) 1.96 eV (632.8 nm) and (b) 2.81 eV (441.6 nm) excitation energies. Raman spectra for 2H (black), 3R (green), and mixed (red) stacking orders are shown. Small circles (●) indicate the shear modes, and the peak marked by ▲ is a higher frequency breathing mode observed only in mixed-phase MoS$_2$. The peaks marked with * is from the silicon substrate.

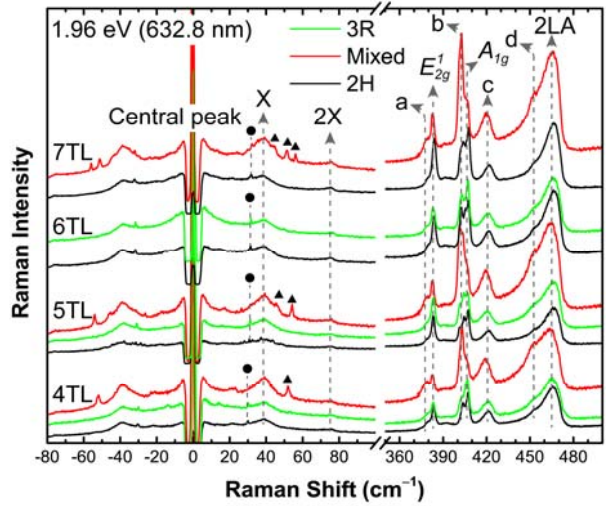

**Figure 4.** Raman spectra of few-layer $MoS_2$ with different stacking orders measured with the excitation energy of 1.96 eV (632.8 nm). Raman spectra for 2H (black), 3R (green), mixed (red) phases are plotted in different colors. Shear modes are marked by small circles (●), and the peaks marked by ▲ are breathing modes observed only in mixed-phase $MoS_2$.

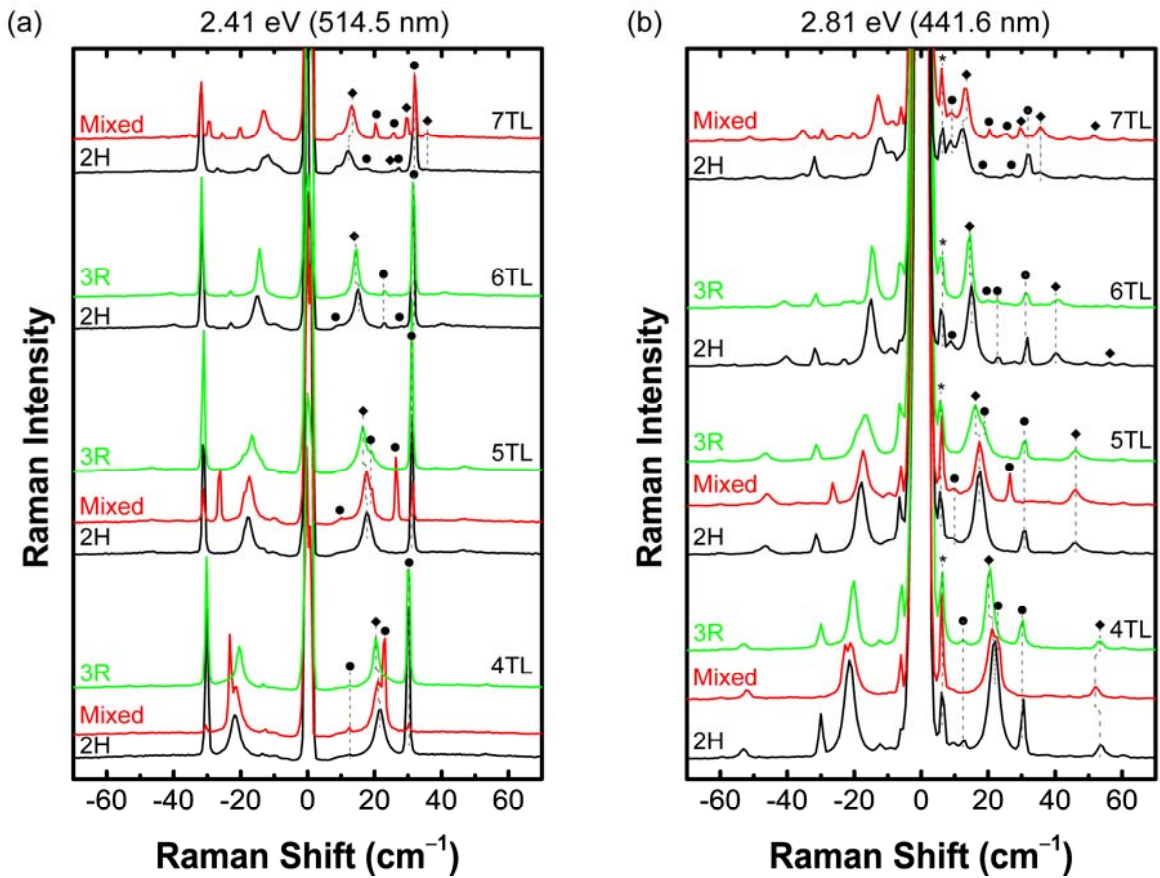

**Figure 5.** Raman spectra of few-layer MoS$_2$ with different stacking orders measured with the excitation energies of (a) 2.41 eV (514.5 nm), and (b) 2.81 eV (441.6 nm). Raman spectra for 2H (black), 3R (green), mixed (red) phases are plotted in different colors. Shear (●) and breathing (♦) modes are marked by symbols. The peak marked by * is from the Si substrate.

ASSOCIATED CONTENT

**Supporting Information.** Crystal structures and space group of polytypes of 3TL-MoS$_2$. HR-TEM images of 2H- and 3R-MoS$_2$, optical microscope images of 3R and mixed phase samples, comparison of the PL spectra, peak positions of Raman modes as a function of the number of TLs with different stacking sequences, Raman spectra of various few-layer MoS$_2$ samples with different stacking sequences. This material is available free of charge *via* the Internet at http://pubs.acs.org.

AUTHOR INFORMATION

**Corresponding Author**

*hcheong@sogang.ac.kr

**Author Contributions**

J.-U.L., K.K., and S.H. prepared the samples. G.H.R. and Z.L. performed HR-TEM measurements. J.-U.L. and K.K. carried out Raman and PL measurements. J.-U.L. and H.C. interpreted the spectroscopic data. The manuscript was written through contributions of all authors. All authors have given approval to the final version of the manuscript.

ACKNOWLEDGEMENT

This work was supported by the National Research Foundation (NRF) grants funded by the Korean government (MSIP) (Nos. 2011-0017605 and 2015R1A2A2A01006992) and by a grant (No. 2011-0031630) from the Center for Advanced Soft Electronics under the Global Frontier Research Program of MSIP.REFERENCES

1. Luo, H.; Xie, W.; Tao, J.; Inoue, H.; Gyenis, A.; Krizan, J. W.; Yazdani, A.; Zhu, Y.; Cava, R. J. Polytypism, Polymorphism, and Superconductivity in TaSe$_{2-x}$Te$_x$. *Proc. Natl. Acad. Sci.* **2015**, *112*, E1174–E1180.

2. Puretzky, A. A.; Liang, L.; Li, X.; Xiao, K.; Wang, K.; Mahjouri-Samani, M.; Basile, L.; Idrobo, J. C.; Sumpter, B. G.; Meunier, V.; Geohegan. D. B. Low-Frequency Raman Fingerprints of Two-Dimensional Metal Dichalcogenide Layer Stacking Configurations. *ACS Nano* **2015**, *9*, 6333–6342.

3. Lu, X.; Utama, M. I. B.; Lin, J.; Luo, X.; Zhao, Y.; Zhang, J.; Pantelides, S. T.; Zhou, W.; Quek, S. Y.; Xiong, Q. Rapid and Nondestructive Identification of Polytypism and Stacking Sequences in Few-Layer Molybdenum Diselenide by Raman Spectroscopy. *Adv. Mater.* **2015**, *27*, 4502–4508.

4. Latil, S.; Henrard, L. Charge Carriers in Few-Layer Graphene Films. *Phys. Rev. Lett.* **2006**, *97*, 036803.

5. Guinea, F.; Castro Neto, A. H.; Peres, N. M. R. Electronic States and Landau Levels in Graphene Stacks. *Phys. Rev. B* **2006**, *73*, 245426.

6. Zou, K.; Zhang, F.; Clapp, C.; MacDonald, A. H.; Zhu, J. Transport Studies of Dual-Gated ABC and ABA Trilayer Graphene: Band Gap Opening and Band Structure Tuning in Very Large Perpendicular Electric Fields. *Nano Lett.* **2013**, *13*, 369–373.

7. Bao, W.; Jing, L.; Velasco, J.; Lee, Y.; Liu, G.; Tran, D.; Standley, B.; Aykol, M.; Cronin, S. B.; Smirnov, D.; Koshino, M.; McCann, E.; Bockrath, M.; Lau, C. N. Stacking-Dependent Band Gap and Quantum Transport in Trilayer Graphene. *Nat. Phys.* **2011**, *7*, 948–952.

8. Lui, C. H.; Li, Z.; Mak, K. F.; Cappelluti, E.; Heinz, T. F. Observation of an Electrically Tunable Band Gap in Trilayer Graphene. *Nat. Phys.* **2011**, *7*, 944–947.
18

Table of Contents Graphic

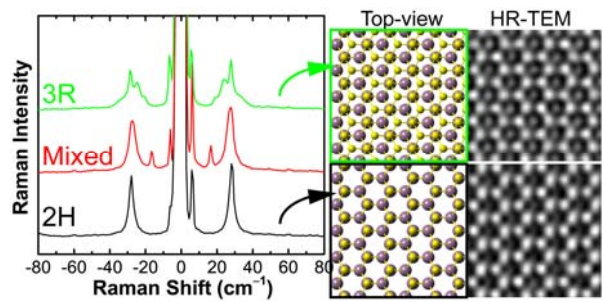